\documentclass[aps,pre,showpacs,twocolumn]{revtex4}
\usepackage{bm}
\usepackage{graphicx}
\bibstyle{approve.bib}
\usepackage{amssymb}
\usepackage{amsmath}
\usepackage{esint}
\usepackage{epstopdf}
\usepackage{color}
\usepackage{gensymb}
\usepackage{mathtools}
\usepackage{suffix}

\newcommand{\be}{\begin{equation}}
\newcommand{\ee}{\end{equation}}
\newcommand{\bea}{\begin{eqnarray}}
\newcommand{\eea}{\end{eqnarray}}
\newcommand{\lan}{\left\langle}
\newcommand{\ran}{\right\rangle}
\newcommand{\br}{\mathbf{r}}

\newcommand{\bE}{\mathbf{E}}

\newcommand{\lm}{L_-}
\newcommand{\lp}{L_+}

\newcommand{\e}{\varepsilon}

\newcommand{\td}{\kappa_b d}
\newcommand{\ta}{\kappa_b a}

\newcommand{\ld}{\lambda_d}
\newcommand{\lb}{\lambda_b}

\begin{document}

\title{Enhanced polymer capture speed and extended translocation time in pressure-solvation traps}

\author{Sahin Buyukdagli$^{1,2}$\footnote{email:~\texttt{buyukdagli@fen.bilkent.edu.tr}}}
\affiliation{$^{1}$Department of Physics, Bilkent University, Ankara 06800, Turkey\\
$^{2}$QTF Centre of Excellence, Department of Applied Physics, Aalto University, FI-00076 Aalto, Finland.}
\date{\small\it \today}

\begin{abstract}

The efficiency of nanopore-based biosequencing techniques requires fast anionic polymer capture by like-charged pores followed by a prolonged translocation process. We show that this condition  can be achieved by setting a pressure-solvation trap. Polyvalent cation addition to the KCl solution triggers the like-charge polymer-pore attraction. The attraction speeds-up the pressure-driven polymer capture but also traps the molecule at the pore exit, reducing the polymer capture time and extending the polymer escape time by several orders of magnitude. By direct comparison with translocation experiments \textcolor{black}{[D. P. Hoogerheide \textit{et al.}, ACS Nano \textbf{8}, 7384 (2014)]}, we \textcolor{black}{characterize} as well the electrohydrodynamics of polymers transport in pressure-voltage traps. \textcolor{black}{We derive scaling laws that can accurately reproduce the pressure dependence of the experimentally measured polymer translocation velocity and time.}  We also find that during polymer capture, the electrostatic barrier on the translocating molecule slows down the liquid flow. This prediction identifies the streaming current measurement as a potential way to probe electrostatic polymer-pore interactions.

\end{abstract}


\date{\today}
\maketitle

\section{Introduction}

The twenty-first century has been witnessing the convergence of previously independent scientific disciplines with the aim of understanding complex structures.  Biopolymer analysis by nanotechnological approaches is a clear example of this scientific turnover~\cite{rev1,rev2}. Along these lines, driven polymer translocation has recently undergone rapid progress~\cite{Tapsarev,e1,e3,e4,e5,e6,e8,e9,e12,exp1,exp2}. Serial sequencing of biopolymers by means of a simple nanopore and an applied voltage offers clear advantages over alternative biosensing techniques that require the biochemical or mechanical modification of each molecule before sequencing.  

The predictive design of polymer translocation devices necessitates primarily the characterization of the electrohydrodynamic and entropic effects governing this highly complex transport process. The entropic contributions from polymer conformations and steric polymer-pore interactions during translocation have been scrutinized by Brownian simulations~\cite{n1,n2,n3,n5} and \textcolor{black}{the tension propagation theory~\cite{Saka1,Saka2,Saka3}}. The electrohydrodynamics of polymer translocation has been considered both by numerical simulations and continuum theories. Monte Carlo (MC) studies by Luan and Aksimentiev investigated the effect of the electroosmotic \textcolor{black}{(EO)} flow~\cite{aks1,aks2} and DNA mobility reversal by polyvalent counterions~\cite{aks3}. By Brownian simulations coupled with a Fokker-Planck (FP) approach, the authors of Ref.~\cite{HLsim} analyzed the electrostatic barrier acting on polymers translocating through $\alpha$-Hemolysin pores. In Ref.~\cite{Chinappi2}, the effect of dipoles placed on the polymer surface was modeled with the aim of extending the translocation time of the molecule. 

Theoretical formulations of \textcolor{black}{purely} voltage-driven polymer transport have been mostly based on mean-field (MF) Poisson-Boltzmann (PB) electrostatics and hydrodynamic Navier-Stokes equation. Along these lines, the poineering drift transport theory developed by Ghosal allowed the consistent derivation of the DNA translocation velocity in terms of the electrophoretic \textcolor{black}{(EP)} and \textcolor{black}{EO} velocity components~\cite{the2,the2II}. Ghosal's mid-pore approximation was subsequently relaxed by Lu et al. via the numerical solution of the coupled PB and Stokes equations~\cite{lu}. The effect of polymer-pore interactions on the unzipping of a DNA hairpin was studied in Ref.~\cite{the3}. Wong and Muthukumar investigated the role played by the \textcolor{black}{EO} flow during diffusion-limited polymer capture by a positively charged pore~\cite{mut1}. Additional models considering the non-equilibrium dynamics of the translocation process upon polymer capture~\cite{mut2,mut3} have been compared with experiments~\cite{mut4}. The details of the polymer hydrodynamics have been also investigated in Refs.~\cite{gros,Rowghanian1,Rowghanian2} by continuum approaches. In Ref.~\cite{the12}, we characterized \textcolor{black}{the correlation-corrected electrohydrodynamics of polymer translocation without the consideration of polymer-pore interactions}. Then in Ref.~\cite{the13}, we incorporated into the electrohydrodynamic transport model of Ref.~\cite{the2II} the repulsive barrier originating from electrostatic polymer-pore interactions \textcolor{black}{at the MF-level}. This improvement extended the drift formalism \textcolor{black}{of Ref.~\cite{the2II}} to include the barrier-limited capture regime prior to translocation. \textcolor{black}{Finally, we have recently extended our purely voltage-driven translocation model of Ref.~\cite{the13} beyond MF level and identified an \textcolor{black}{electroosmotically} facilitated polymer capture mechanism~\cite{theS}.}

\textcolor{black}{Polymers can alternatively be transported by an externally applied hydrostatic pressure gradient between the cis and trans sides of the membrane. The pressure gradient induces a streaming flow through the pore. The drag force exerted by this streaming current carries the polymer from the cis to trans side of the membrane.} At the theoretical level, \textcolor{black}{streaming flow-driven} polymer transport has received less attention than its \textcolor{black}{electrohydrodynamic} counterpart. Solving Edward's polymer diffusion equation, Stein et al. studied entropic effects on polymer transport through nanoslits~\cite{Stein}. In Ref.~\cite{Buyuk2015}, we predicted ionic correlation-induced streaming current inversion in pressure-driven polymer translocation events. At this point, we note that the precision of polymer translocation requires, among other factors, the extension of the translocation time upon polymer capture~\cite{e5}.  \textcolor{black}{Translocation experiments by Hoogerheide et al. showed that this goal can be achieved by setting a pressure-voltage trap, which consists of imposing a pressure gradient with the aim of counterbalancing the external voltage~\cite{exp1,exp2}. It was observed that the resulting suppression of the net drift force allows to trap the translocating molecule without causing significant perturbation of the ionic current signal.} Via the numerical solution of the electrohydrodynamic formalism of Ref.~\cite{lu} coupled with an effective diffusion equation, the experimental data of translocation time was also interpreted in Ref.~\cite{exp2}. 

\textcolor{black}{In this article, we characterize the additional effect of direct electrostatic polymer-membrane interactions in polymer translocation events driven by a pressure and a voltage. To this end}, in Section~\ref{mod}, we extend the voltage-driven transport model of Ref.~\cite{the13} to include the streaming current induced by an applied pressure gradient. Section~\ref{mf} deals with the electrohydrodynamic mechanism driving such a pressure-voltage trap.  \textcolor{black}{First, we confront our theory with the experiments of Ref.~\cite{exp2}. We show that our newly derived  scaling laws~(\ref{vm}) and~(\ref{tausc}) can quantitatively describe the experimentally measured evolution of the polymer translocation velocity and time with the pressure gradient.} Then, in terms of the experimentally tunable system parameters, we fully characterize the polymer conductivity of anionic pores under pressure-voltage traps. Our theory also predicts that during polymer capture, like-charge polymer-pore interactions transmitted to the liquid by the drag force slow down the liquid flow. This suggests that the nature and magnitude of electrostatic polymer-pore interactions can be extracted from streaming current measurements.

In addition to a prolonged polymer translocation, the efficiency of nanopore-based sequencing methods requires  fast polymer capture by the pore. Considering that most of the silicon-based solid-state pores carry negative surface charges of high density~\cite{e6}, the technical challenge consists in driving as fast as possible an anionic polymer into a like-charged pore by overcoming the electrostatic polymer-pore repulsion. In Section~\ref{cr}, we show that rapid polymer capture and extended translocation can be mutually achieved by setting a pressure-solvation trap driven by charge correlations. To this end, we generalize the formulation of polymer pore-interactions beyond-MF level. This extension is introduced within the test charge theory of Ref.~\cite{the14} explained in Section~\ref{barcr}.  We note that the test charge theory has been previously shown to accurately describe the experimentally observed similar charge attraction between polyelectrolytes~\cite{exp4,the11} and polymer-membrane complexes~\cite{Molina2014,the110}. 

Within this correlation-corrected pressure-driven transport formalism, we show that polyvalent cations added to the KCl solution amplify electrostatic correlations and turn polymer-pore interactions from repulsive to attractive. This like-charge attraction enhances the polymer capture speed but also traps the molecule at the pore exit, reducing the barrier-limited polymer capture time and extending the polymer escape time by several orders of magnitude. This result is the key prediction of our work. We note that a similar trapping mechanism resulting from the inversion of the fixed pore charge upon pH variation has been experimentally observed in translocation events in $\alpha$-hemolysin pores~\cite{mutalp}. In terms of the experimentally controllable system parameters, we throughly identify the parameter regime maximizing the enhancement of the polymer capture speed and escape time \textcolor{black}{by the electrostatic trap. It should be noted that this  trapping mechanism differs from the facilitated polymer capture process of Ref.~\cite{theS} where the polymer capture speed is enhanced by the EO flow rather than polymer-pore interactions.} The approximations and possible improvements of our model are elaborated in Conclusions.

\section{Translocation Model}
\label{mod}

\begin{figure}
\includegraphics[width=1.\linewidth]{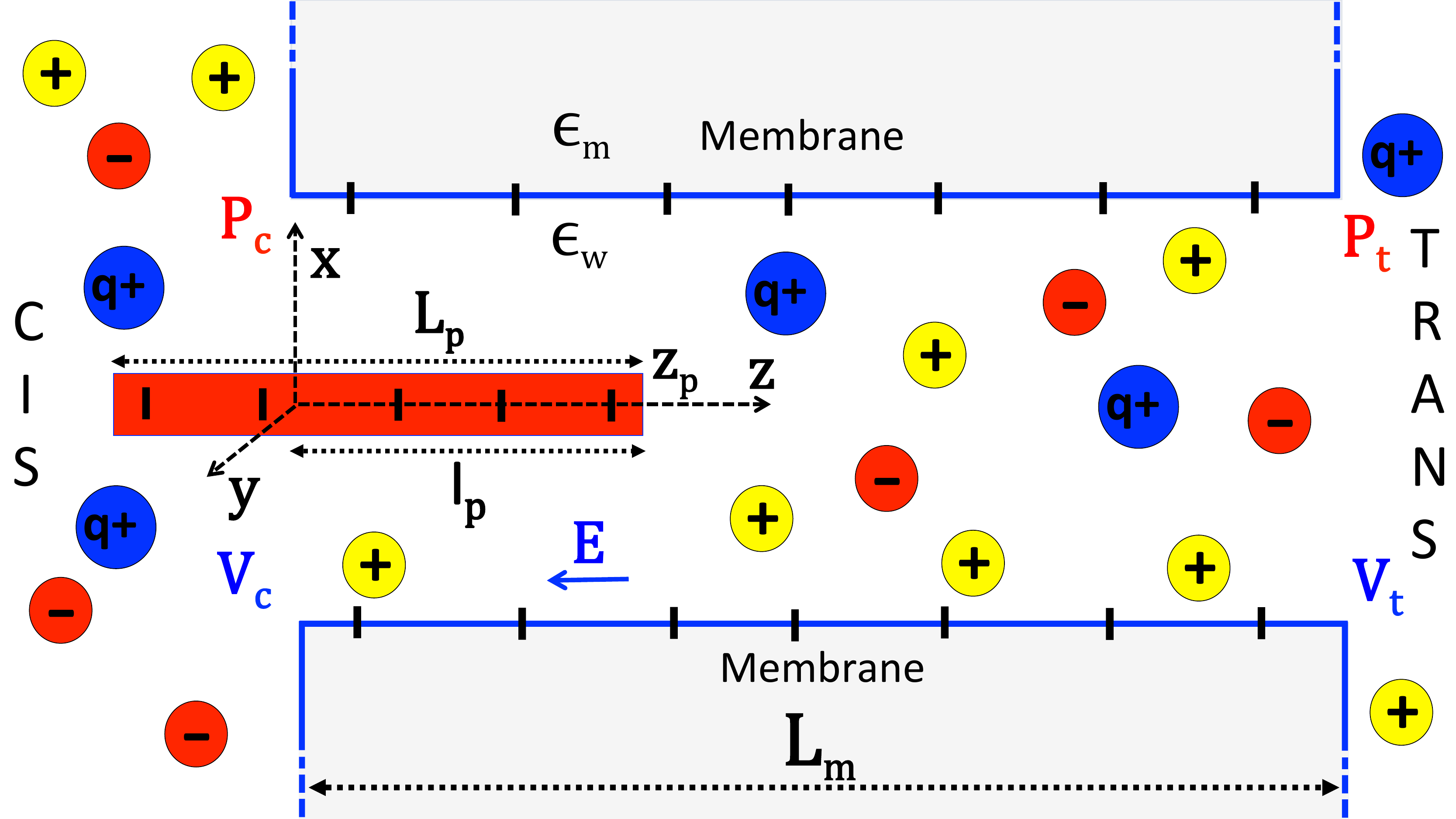}
\caption{(Color online) Schematic depiction of the pore with length $L_m$, radius $d$, and negative wall charge density $-\sigma_m$. The confined solution includes monovalent $\mbox{K}^+$ and $\mbox{Cl}^-$ ions, and multivalent cations of valency $q$. The dielectric permittivities of the pore and the membrane are $\e_w=80$ and $\e_m=2$. The polymer of length $L_p$, radius $a$, charge density $-\sigma_p$,  and the right end position $z_p$ translocates under the effect of the pressure gradient $\Delta P=P_c-P_t$ and voltage $\Delta V=V_t-V_c$. The electric field $\bE=-E\hat{u}_z$ has magnitude $E=\Delta V/L_m$.}
\label{fig1}
\end{figure}

Our translocation model is depicted in Fig.~\ref{fig1}. The cylindrical nanopore of radius $d$, length $L_m$, and negative surface charge density $-\sigma_m$ is in contact with a reservoir containing the KCl electrolyte,  a multivalent cation species of valency $q>0$, and \textcolor{black}{anionic polymers of low concentration whose interactions can be neglected}. The reservoir concentration of the ionic species $i$ is $\rho_{bi}$, and the bulk electroneutrality reads $\rho_{b+}-\rho_{b-}+q\rho_{bq+}=0$. The dielectric permittivities of the pore and the membrane are respectively $\e_w=80$ and $\e_m=2$. \textcolor{black}{Considering that dsDNA has a large persistence length of about $50$ nm, we neglect conformational polymer fluctuations. Thus, the translocating polymer is modelled as a rigid cylinder of length $L_p$ and typical radius $a=1$ nm of dsDNA molecules. The discrete helicoidal charge distribution on the DNA backbone is approximated by a continuous surface charge density $-\sigma_p$, with the numerical value $\sigma_p=0.4$ $e/\mbox{nm}^2$ previously obtained by fitting experimental current blockage data~\cite{the12}. Polymer translocation from the cis to trans side occurs} under the effect of the applied voltage $\Delta V$ and pressure $\Delta P$, and the potential barrier $V_p(z_p)$ resulting from electrostatic polymer-membrane interactions.

The translocation dynamics is characterized by the \textcolor{black}{polymer number density} $c(z_p,t)$ satisfying the Smoluchowski equation~\cite{muthu,mut2}
\bea\label{con}
\partial_tc(z_p,t)&=&-\partial_{z_p}J(z_p,t)\\
\label{cur}
J(z_p,t)&=&-D\partial_{z_p}c(z_p,t)+v_p(z_p)c(z_p,t),
\eea
where $z_p$ is the position of the polymer with diffusion coefficient $D=\ln(L_p/2a)/(3\pi\eta L_p\beta)$~\cite{cyl1,cyl2}, with the inverse thermal energy $\beta=1/(k_BT)$, the Boltzmann constant $k_B$, the liquid temperature $T=300$ K, and the solvent viscosity $\eta=8.91\times10^{-4}$ Pa s.  \textcolor{black}{Furthermore, $J(z_p,t)$ stands for the net polymer flux through the pore, with the} polymer velocity  
\be\label{vp}
v_p(z_p)=-\beta DU'_p(z_p)
\ee
where $U_p(z_p)$ is the polymer potential that will be derived below. At steady state  with constant \textcolor{black}{polymer density}, $\partial_tc(z_p,t)=0$, the integration of the uniform \textcolor{black}{flux} condition $J(z_p,t)=J_0$ together with the fixed density condition at the pore entrance $c(z_p=0)=c_{cis}$ and an absorbing boundary at the pore exit $c(z_p=L_p+L_m)=0$ yields  the polymer \textcolor{black}{number} density in the form
\be
\label{polden}
c(z_p)=c_{cis}e^{-\beta U_p(z_p)}\frac{\int_{z_p}^{L_p+L_m}\mathrm{d}z\;e^{\beta U_p(z)}}{\int_{0}^{L_p+L_m}\mathrm{d}z\;e^{\beta U_p(z)}}.
\ee
Moreover, the translocation rate defined as the polymer current per density $R_c\equiv J_0/c_{cis}$ follows as
\be
\label{rc}
R_c=\frac{D}{\int_0^{L_m+L_p}\mathrm{d}ze^{\beta U_p(z_p)}}.
\ee
\textcolor{black}{We finally note that in the dilute polymer regime where polymer interactions are negligible, the number density~(\ref{polden}) is equivalent to the polymer probability function.}

The following part generalizes the electrohydrodynamic transport model of Ref.~\cite{the13} to include the pressure gradient. In order to derive the polymer potential $U_p(z_p)$, we introduce first the PB and Stokes equations for  the electrostatic potential $\phi(r)$ and convective fluid velocity $u_c(r)$  in the pore,
\bea\label{pb}
&&\frac{1}{r}\partial_r\left[r\partial_r\phi(r)\right]+4\pi\ell_B\left[\rho_c(r)+\sigma(r)\right]=0\\
\label{st}
&&\frac{\eta}{r}\partial_r\left[r\partial_ru_c(r)\right]-e\rho_c(r)E+\frac{\Delta P}{L_m}=0,
\eea 
with the radial distance $r$ from the pore axis, the Bjerrum length $l_B=\beta e^2/(4\pi\e_w)$, the electron charge $e$, and the density of mobile charges $\rho_c(r)=\sum_{i=1}^3q_i\rho_{bi}e^{-q_i\phi(r)}$ and fixe charges $\sigma(r)=-\sigma_m\delta(r-d)-\sigma_p\delta(r-a)$. \textcolor{black}{In Eqs.~(\ref{pb}) and~(\ref{st}), the cylindrical symmetry of the model was preserved by neglecting electrohydrodynamic edge effects associated with the finite pore length. This approximation is justified by the fact that the pore and polymer lengths considered in our work are much larger than the Bjerrum length $\ell_B\approx7$ {\AA} corresponding to the spatial scale where  finite electrohydrodynamic size effects on polymer capture would be relevant. In Sec.~\ref{exp}, this point will be confirmed by comparison with experiments. Now,} we combine the PB and Stokes Eqs.~(\ref{pb})-(\ref{st}) to eliminate the density $\rho_c(r)$, and integrate the result with the no-slip boundary condition at the pore wall $u_c(d)=0$ and at the DNA surface $u_c(a)=v_p(z_p)$. Finally, we account for Gauss' law $\phi'(a)=4\pi\ell_B\sigma_p$ and the force balance relation on the polymer $F_{el}+F_{dr}+F_b=0$, with the electrostatic force $F_{el}=2\pi aL_peE$, the drag force $F_{dr}=2\pi aL_p\eta u'_c(a)$, and the barrier-induced force $F_b=-V'_p(z_p)$. After some algebra, the liquid and polymer velocities follow as
\bea\label{vels}
u_c(r)&=&\mu_eE\left[\phi(d)-\phi(r)\right]-\beta D_p(r)\frac{\partial V_p(z_p)}{\partial z_p}\nonumber\\
&&+\frac{\Delta P}{4\eta L_m}\left[d^2-r^2-2a^2\ln\left(\frac{d}{r}\right)\right]\\
\label{velp}
v_p(z_p)&=&v_{dr}-\beta D_p(a)\frac{\partial V_p(z_p)}{\partial z_p},
\eea
with the effective diffusion coefficient in the pore
\be\label{efd}
D_p(r)=\frac{\ln(d/r)}{2\pi\eta L_p\beta},
\ee
\textcolor{black}{EP} mobility coefficient $\mu_e=\e_w k_BT/(e\eta)$, and the drift velocity component
\be
\label{vdr}
v_{dr}=\frac{\mu_e\Delta V}{L_m}\left[\phi(d)-\phi(a)\right]+\frac{\gamma a^2\Delta P}{4\eta L_m},
\ee
where
\be\label{gm}
\gamma=\frac{d^2}{a^2}-1-2\ln\left(\frac{d}{a}\right). 
\ee
Combining Eqs.~(\ref{vp}) and~(\ref{velp}), and integrating the result, the effective polymer potential that determines the density~(\ref{polden}) finally becomes
\be
\label{polp}
U_p(z_p)=\frac{D_p(a)}{D}V_p(z_p)-\frac{v_{dr}}{\beta D}z_p.
\ee

In Eq.~(\ref{polp}), the interaction potential corresponds to the electrostatic coupling energy between the fixed pore and polymer charges, 
\be\label{vp1}
V_p(z_p)=\Delta\Omega_p\left[l_p(z_p)\right],
\ee
where $\Delta\Omega_p(l_p)$ stands for the electrostatic grand potential of the polymer portion located in the pore. The position-dependent length of this portion reads
\bea
\label{lpzp}
l_p(z_p)&=&z_p\theta(L_--z_p)+L_-\theta(z_p-L_-)\theta(L_+-z_p)\nonumber\\
&&+(L_p+L_m-z_p)\theta(z_p-L_+),
\eea
with the auxiliary lengths 
\be\label{lpm}
L_-=\mathrm{min}(L_m,L_p);\hspace{5mm}L_+=\mathrm{max}(L_m,L_p). 
\ee
The explicit form of the polymer grand potential $\Delta\Omega_p(l_p)$ in Eq.~(\ref{vp1}) will be specified in Sections~\ref{mf} and~\ref{cr} according to the approximation level.

\section{Pressure-voltage traps}
\label{mf}

We characterize here the pressure-voltage-driven translocation of polymers in the monovalent KCl solution of reservoir concentration $\rho_b$. Electrostatic correlations being negligible in monovalent electrolytes, charge interactions will be formulated within MF electrostatics.

\subsection{Computation of the drift velocity and electrostatic barrier}
\label{elmf}

According to Eq.~(\ref{vdr}), the computation of the drift velocity $v_{dr}$ in Eq.~(\ref{polp}) requires the knowledge of the pore potential $\phi(r)$. In the cylindrical pore geometry, the corresponding PB Eq.~(\ref{pb}) does not possess a closed-form solution. Within an improved Donnan approximation that allows to preserve the non-linearity of Eq.~(\ref{pb}), the pore  potential was derived in Ref.~\cite{the13} in the form
\bea
\label{pot1}
\phi(r)&=&-\ln\left(t+\sqrt{t^2+1}\right)+\frac{8\pi\ell_B}{\kappa_d^2}\frac{\sigma_md+\sigma_pa}{d^2-a^2}\\
&&+\frac{4\pi\ell_B}{\kappa_d}\frac{T_1\mathrm{I}_0(\kappa_dr)+T_2\mathrm{K}_0(\kappa_dr)}{\mathrm{I}_1(\kappa_da)\mathrm{K}_1(\kappa_dd)-\mathrm{K}_1(\kappa_da)\mathrm{I}_1(\kappa_dd)}\nonumber.
\eea
In Eq.~(\ref{pot1}), we introduced the ratio of the membrane and pore charge densities $t=(d\sigma_m+a\sigma_p)/[\rho_b(d^2-a^2)]$, the auxiliary coefficients $T_1=\sigma_m\mathrm{K}_1(\kappa_da)+\sigma_p\mathrm{K}_1(\kappa_dd)$ and $T_2=\sigma_m\mathrm{I}_1(\kappa_da)+\sigma_p\mathrm{I}_1(\kappa_dd)$ with the modified Bessel functions $\mathrm{I}_m(x)$ and $\mathrm{K}_m(x)$~\cite{math}, and the effective pore screening and bare Debye-H\"{u}ckel parameters 
\be
\kappa_d=\kappa_b\left(1+t^2\right)^{1/4}\hspace{0mm};\hspace{5mm}\kappa_b=\sqrt{8\pi\ell_B\rho_b}.
\ee
Inserting the potential~(\ref{pot1}) into Eq.~(\ref{vdr}), the drift velocity becomes
\be\label{vdr2}
v_{dr}=\frac{4\pi\ell_B\mu_e\Theta\Delta V}{\kappa_dL_m}+\frac{\gamma a^2\Delta P}{4\eta L_m},
\ee
with the auxiliary coefficient
\be
\Theta=\frac{T_1\left[\mathrm{I}_0(\kappa_dd)-\mathrm{I}_0(\kappa_da)\right]+T_2\left[\mathrm{K}_0(\kappa_dd)-\mathrm{K}_0(\kappa_da)\right]}{\mathrm{I}_1(\kappa_da)\mathrm{K}_1(\kappa_dd)-\mathrm{K}_1(\kappa_da)\mathrm{I}_1(\kappa_dd)}.
\ee

The MF level interaction energy between the polymer portion in the pore and the fixed pore charges reads
\be\label{grmf}
\beta\Delta\Omega_p(l_p)=\int\mathrm{d}\br\sigma_p(\br)\phi_m(\br). 
\ee
The polymer charge density is 
\be\label{chp}
\sigma_p(\br)=-\sigma_p\delta(r-a)\theta(z)\theta(l_p-z). 
\ee
The electrostatic potential $\phi_m(r)$ induced exclusively by the pore charges follows from Eq.~(\ref{pot1}) by setting $\sigma_p=0$,
\bea
\label{pot2}
\phi_m(r)&=&-\ln\left(t_m+\sqrt{t_m^2+1}\right)+\frac{4}{\mu_m\kappa_m^2}\frac{d}{d^2-a^2}\\
&&+\frac{2}{\mu_m\kappa_m}\frac{\mathrm{K}_1(\kappa_{m}a)\mathrm{I}_0(\kappa_mr)+\mathrm{I}_1(\kappa_ma)\mathrm{K}_0(\kappa_mr)}{\mathrm{I}_1(\kappa_ma)\mathrm{K}_1(\kappa_md)-\mathrm{K}_1(\kappa_ma)\mathrm{I}_1(\kappa_md)},\nonumber
\eea
with the charge ratio $t_m=d\sigma_m/[\rho_b(d^2-a^2)]$, the screening parameter $\kappa_m=\kappa_b\left(1+t_m^2\right)^{1/4}$, and the Gouy-Chapman length $\mu_m=1/(2\pi\ell_B\sigma_m)$. Substituting the charge density~(\ref{chp}) into Eq.~(\ref{grmf}), the interaction potential~(\ref{vp1}) finally becomes 
\be\label{Vp}
V_p(z_p)=-2\pi a\sigma_pk_BT\phi_m(a)l_p(z_p).
\ee
In an anionic pore where $\phi_m(a)<0$, the potential~(\ref{Vp}) rises with the penetration length $l_p$. Thus, this potential acts as an electrostatic barrier that limits the polymer capture. Finally, introducing the characteristic inverse lengths associated with the drift~(\ref{vdr2}) and the barrier~(\ref{Vp}),
\be
\label{25}
\ld=\frac{v_{dr}}{D};\hspace{5mm}\lb=-2\pi a\sigma_p\phi_m(a)\frac{D_p(a)}{D},
\ee
the polymer velocity~(\ref{velp}) and potential~(\ref{polp}) follow as
\bea\label{vp2}
v_p(z_p)&=&v_{dr}-D\lambda_b\left[\theta(L_--z_p)-\theta(z_p-L_+)\right]\\
\label{up2}
\beta U_p(z_p)&=&\lb l_p(z_p)-\ld z_p.
\eea

\subsection{Comparison with trapping experiments}
\label{exp}

Using the polymer density function~(\ref{polden}) and Eqs.~(\ref{vp2})-(\ref{up2}), we calculate first the average polymer velocity 
\be\label{bv0}
\lan v_p\ran=\frac{\int_0^{L_p+L_m}\mathrm{d}z_p c(z_p)v_p(z_p)}{\int_0^{L_p+L_m}\mathrm{d}z_p c(z_p)}.
\ee
Carrying out the integrals in Eq.~(\ref{bv0}), one obtains
\bea\label{bv}
\lan v_p\ran=v_{dr}-D\lambda_b\frac{J_1-J_3}{J_1+J_2+J_3},
\eea
where the coefficients $J_{i=1,2,3}$ depending on the parameters $\lambda_{d,b}$ and $L_\pm$ are reported in Appendix~\ref{coef}. In Fig.~\ref{fig2}(a), we display the pressure dependence of the velocity~(\ref{bv}) together with the experimental velocity data of Ref.~\cite{exp2}. The experimental parameters taken from Ref.~\cite{exp2} are the voltage $\Delta V=-100$ mV, the salt density $\rho_b=1.6$ M, the monomer number $N=615$ bps corresponding to the polymer length $L_p=180$ nm, and the pore radius $d=5$ nm. The pore length and charge density were adjusted to the values $L_m=200$ nm~\cite{rem1} and $\sigma_m=0.13$ $e/\mbox{nm}^2$  that provided the best agreement with the magnitude of the velocity data. The charge density value is comparable with the experimental value $\sim30$ $\mbox{mC}/\mbox{m}^2\approx0.18$ $e/\mbox{nm}^2$ measured at the solution $\mbox{pH}\sim8$~\cite{Hooger} where the translocation experiments of Ref.~\cite{exp2} were carried-out. 

In the barrier-driven regime $\lb\gg\ld$, Eq.~(\ref{bv}) simplifies to $\lan v_p\ran\approx D(\ld-\lb)$. Passing to the linear PB approximation, and expanding the inverse lengths of Eq.~(\ref{25}) in terms of $\sigma_p$ and $\sigma_m$,  the velocity follows as
\be
\label{bvln}
\lan v_p\ran\approx\frac{f_p\sigma_p-f_m\sigma_m}{g\kappa_b\eta}\frac{e\Delta V}{L_m}+\frac{\gamma a^2\Delta P}{4\eta L_m}-\frac{e^2\sigma_p\sigma_m\ln(d/a)}{g\eta\e_w\kappa_b^2L_p},
\ee
where we introduced the geometric coefficients 
\bea
\label{fg}
f_p&=&\mathrm{K}_1(\td)\mathrm{I}_0(\ta)+\mathrm{I}_1(\td)\mathrm{K}_0(\ta)-\left(\td\right)^{-1}\\
f_m&=&\mathrm{K}_1(\ta)\mathrm{I}_0(\td)+\mathrm{I}_1(\ta)\mathrm{K}_0(\td)-\left(\ta\right)^{-1}\\
g&=&\mathrm{I}_1(\td)\mathrm{K}_1(\ta)-\mathrm{I}_1(\ta)\mathrm{K}_1(\td).
\eea
The approximation~(\ref{bvln}) derived in the barrier-dominated regime will be shown to work as well in the drift-driven  regime $\lb\ll\ld$ where  $\lan v_p\ran\approx D\ld\approx D(\ld-\lb)$.

\begin{figure}
\includegraphics[width=.9\linewidth]{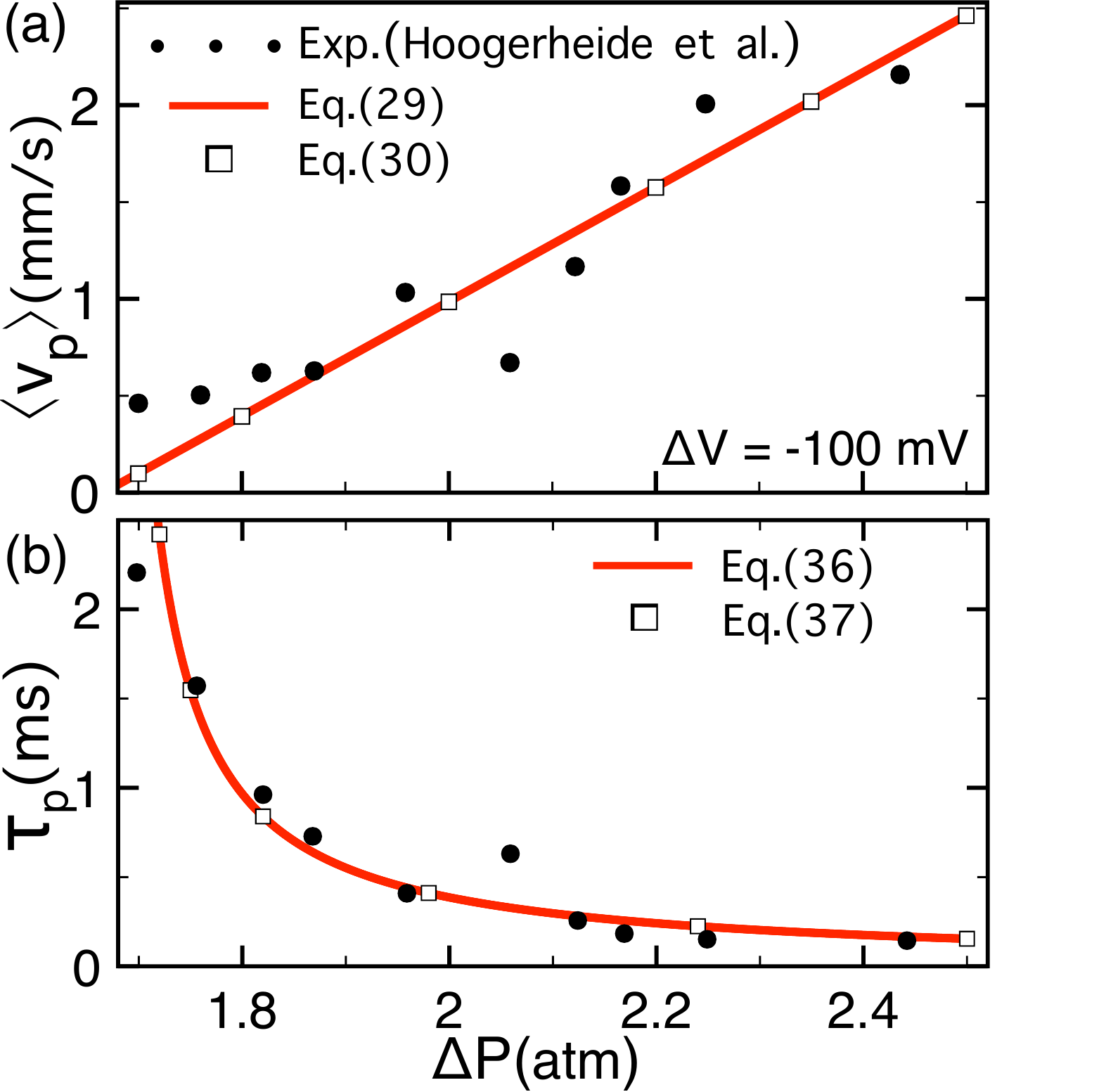}
\caption{(Color online) (a) Average polymer velocity $\lan v_p\ran$ and (b) translocation time $\tau_p=(L_m+L_p)/\lan v_p\ran$ versus pressure. Solid curves are from Eq.~(\ref{bv}) and squares mark the linear result~(\ref{bvln}). The experimental velocity data in (a) are from Fig.S3 of the supporting information of Ref.~\cite{exp2}. The data of average escape time in (b) are from Fig.4(b) of Ref.~\cite{exp2}. The model parameters are given in the main text.}
\label{fig2}
\end{figure}

The first component of Eq.~(\ref{bvln}) accounts for the \textcolor{black}{EP} drift (positive term) and the \textcolor{black}{EO} drag (negative term). The second and third components originate respectively from the streaming current, and the electrostatic barrier induced by like-charge polymer-membrane repulsion that hinders the polymer capture. Eq.~(\ref{bvln}) reported in Fig.~\ref{fig2}(a) indicates that as a result of the drag force induced by the streaming flow, the average velocity rises linearly with pressure as
\be\label{vm}
\lan v_p\ran\approx\frac{\gamma a^2}{4\eta L_m}\left(\Delta P-\Delta P^*\right),
\ee
with the critical pressure for polymer trapping
\be
\label{prcr}
\Delta P^*=-\frac{4\left(f_p\sigma_p-f_m\sigma_m\right)}{\gamma ga^2\kappa_b}e\Delta V +\frac{4\ln(d/a)e^2\sigma_p\sigma_mL_m}{\gamma ga^2\e_w\kappa_b^2L_p}.
\ee

A successful translocation requires the polymer to travel the distance $L_m+L_p$. The translocation time can thus be estimated in terms of the velocity~(\ref{bv}) as  
\be\label{tau1}
\tau_p\approx\frac{L_m+L_p}{\lan v_p\ran}.
\ee
Fig.~\ref{fig2}(b) shows that with the same parameters as in Fig.~\ref{fig2}(a), this theoretical estimation can accurately reproduce the experimental escape times of Ref.~\cite{exp2}. The linear PB approximation for $\tau_p$ obtained from Eq.~(\ref{vm}) 
\be\label{tausc}
\tau_p\approx\frac{4\eta L_m\left(L_p+L_m\right)}{\gamma a^2\left(\Delta P-\Delta P^*\right)}
\ee
indicates that the quick rise of the experimental escape time with decreasing pressure occurs according to an inverse power law (see the square symbols). 

\begin{figure}
\includegraphics[width=1.0\linewidth]{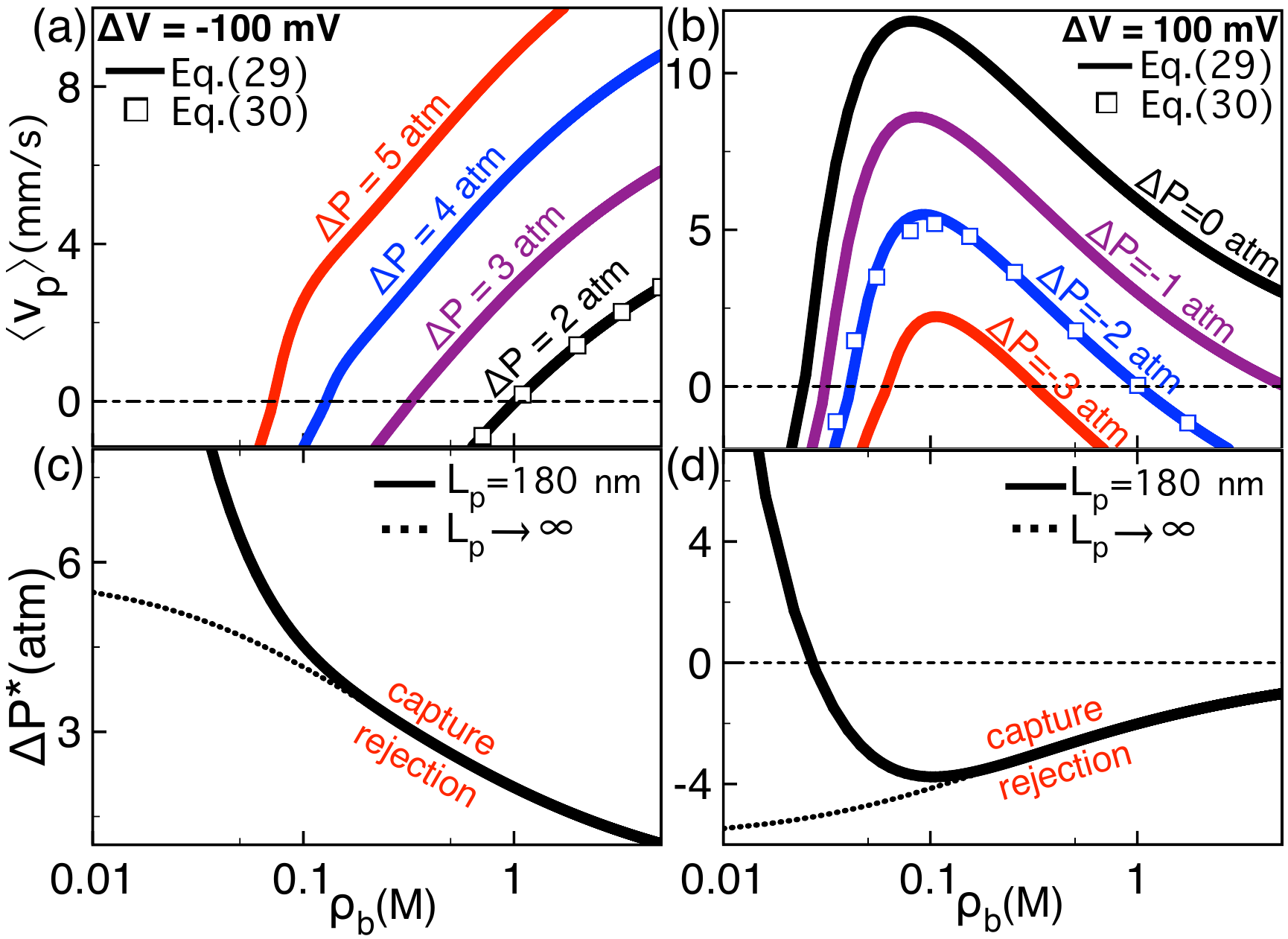}
\caption{(Color online) (a)-(b)  Salt dependence of the average polymer velocity~(\ref{bv}) at various pressure gradients. (c)-(d) The critical pressure gradient~(\ref{prcr}) for polymer trapping. The voltage is $\Delta V=-100$ mV (left plots) and $100$ mV (right plots). The other parameters are the same as in Fig.~\ref{fig2}.}
\label{fig3}
\end{figure}

\subsection{Effect of salt, polymer length, and pore size}
\label{fin}

We scrutinize here the effect of the experimentally tuneable parameters on polymer trapping. Figs.~\ref{fig3}(a) and (b) illustrate the salt dependence of the polymer velocity and also show the accuracy of the approximation~(\ref{bvln}) (square symbols). In Fig.~\ref{fig3}(a) where translocation is driven by the streaming current ($\Delta P>0$) and limited by voltage ($\Delta V<0$), the increment of the ion density rises the polymer velocity ($\rho_b\uparrow\lan v_p\ran\uparrow$) and switches its sign from negative to positive. Thus, added salt favours polymer capture. In order to gain analytical insight into this effect, we expand Eq.~(\ref{bvln}) in the corresponding strong salt regime $\kappa a\gg1$ and $\kappa d\gg1$ to obtain
\be\label{dil1}
\lan v_p\ran\approx\frac{(\sigma_p-\sigma_m)e\Delta V}{\eta L_m\kappa_b}+\frac{\gamma a^2\Delta P}{4\eta L_m}.
\ee 
According to Eq.~(\ref{dil1}), the velocity increase by added salt originates from the screening of the voltage-induced drift opposing the polymer capture. Due to the same screening effect, in Fig.~\ref{fig3}(b) where polymer transport is driven by voltage ($\Delta V>0$), added salt of high density ($\rho_b\gtrsim0.1$ M) turns the velocity from positive to negative ($\rho_b\uparrow\lan v_p\ran\downarrow$) and blocks polymer transport. Setting Eq.~(\ref{dil1}) to zero, the ion concentration for polymer trapping in strong salt follows as
\be
\label{dil2}
\rho_{b>}\approx\frac{2}{\pi\ell_B}\left[\frac{(\sigma_p-\sigma_m)e\Delta V}{\gamma a^2\Delta P}\right]^2.
\ee
In agreement with Figs.~\ref{fig3}(a) and (b), Eq.~(\ref{dil2}) predicts the reduction of the characteristic salt density with increasing pressure gradient, i.e. $|\Delta P|\uparrow\rho_{b>}\downarrow$.  

\begin{figure*}
\includegraphics[width=.95\linewidth]{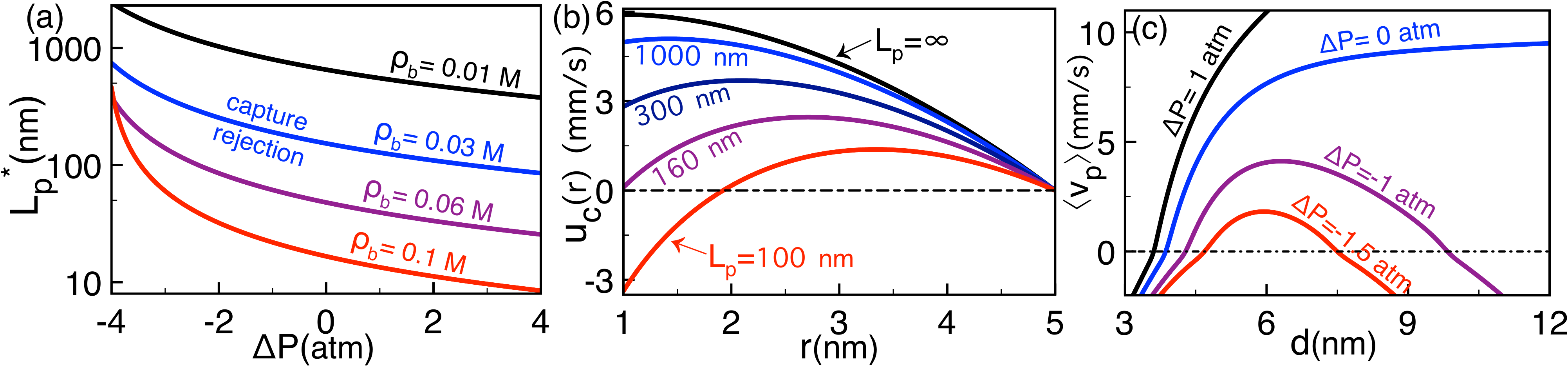}
\caption{(Color online) (a) Critical polymer length~(\ref{lcr}) against the pressure $\Delta P$ at the voltage $\Delta V=100$ mV. (b) Liquid velocity~(\ref{vels}) at vanishing voltage $\Delta V=0$ and  pressure $\Delta P=2$ atm. (c) Polymer velocity~(\ref{bv}) against the pore radius at the voltage $\Delta V=100$ mV. The salt density in (b) and (c) is $\rho_b=0.05$ M. The other parameters are the same as in Fig.~\ref{fig2}.}
\label{fig4}
\end{figure*}

In the dilute salt regime of Fig.~\ref{fig3}(b), one notes the presence of a second critical salt density where the velocity cancels. To explain the origin of this reversal point, we expand Eq.~(\ref{bvln}) for $\kappa a\ll1$ and $\kappa d\ll1$ to get
\bea
\label{dil3}
\lan v_p\ran&\approx&\frac{(a_p\sigma_p-a_m\sigma_m)e\Delta V}{\eta L_m}+\frac{\gamma a^2\Delta P}{4\eta L_m}\\
&&-\frac{da\ln(d/a)}{d^2-a^2}\frac{k_BT\sigma_p\sigma_m}{\eta L_p\rho_b},\nonumber
\eea
with the auxiliary coefficients 
\be\label{apm}
a_p=-\frac{a}{2}+\frac{ad^2\ln(d/a)}{d^2-a^2};\hspace{3mm}a_m=\frac{d}{2}-\frac{a^2d\ln(d/a)}{d^2-a^2}.
\ee
Eq.~(\ref{dil3}) indicates that in Fig.~\ref{fig3}(b), enhanced polymer conductivity by added salt ($\rho_b\uparrow\lan v_p\ran\uparrow$) stems from the screening of repulsive polymer-membrane interactions. Thus, polymer trapping at dilute salt originates from the competition between the drift force and the electrostatic barrier. The corresponding salt concentration follows from Eq.~(\ref{dil3})  as
\be\label{dil4}
\rho_{b<}\approx\frac{4da\ln(d/a)L_m}{(d^2-a^2)L_p}\frac{k_BT\sigma_p\sigma_m}{\gamma a^2\Delta P+4(a_p\sigma_p-a_m\sigma_m)e\Delta V}.
\ee
In accordance with Fig.~\ref{fig3}(b), Eq.~(\ref{dil4}) predicts the rise of the lower critical salt concentration by enhanced negative pressure, i.e. $|\Delta P|\uparrow \rho_{b<}\uparrow$. 

The phase diagrams of Figs.~\ref{fig3}(c) and (d) illustrate the salt dependence of the critical pressure~(\ref{prcr}). One sees that regardless of the voltage sign, the critical pressure is reduced by dilute salt, i.e.  $\rho_b\uparrow\Delta P^*\downarrow$. The low ion density expansion of Eq. (35) 
\be
\label{prdl}
\Delta P^*\approx-\frac{4\left(a_p\sigma_p-a_m\sigma_m\right)}{\gamma a^2}e\Delta V+\frac{4da\ln(d/a)\sigma_p\sigma_mL_m}{\beta\gamma a^2(d^2-a^2)L_p\rho_b}.
\ee
indicates that this behavior results from the screening of the electrostatic barrier. In voltage-driven transport ($\Delta V>0$), this trend is reversed in the strong salt regime where the critical pressure rises, $\rho_b\uparrow\Delta P^*\uparrow$.  The high density expansion of Eq.~(35)
\be\label{str1}
\Delta P^*\approx-\frac{4(\sigma_p-\sigma_m)e\Delta V}{\gamma a^2\kappa_b}
\ee
shows that the rise of $\Delta P^*$ is due to the shielding of the voltage-induced drift force on DNA.

We consider now the effect of the finite polymer length. According to Eq.~(\ref{prdl}), in the dilute salt regime, the capture of shorter polymers requires higher pressures, i.e. $L_p\downarrow \Delta P^*\uparrow$. This finite-size effect is also displayed in Figs.~\ref{fig3}(c) and (d). The obstruction of polymer capture by finite molecular length is due to the repulsive barrier term of Eq.~(\ref{bvln}); the streaming current and voltage act on the whole polymer of length $L_p$  while the barrier affects solely the polymer portion in the pore. Hence, the net drag force on the polymer decreases with the length of the molecule. As a result, the polymer velocity~(\ref{bvln}) drops with decreasing polymer length ($\L_p\downarrow \lan v_p\ran\downarrow$) as  
\be\label{l1}
\lan v_p\ran\approx v_{dr}\left(1-\frac{L^*_p}{L_p}\right),
\ee
with the critical molecular length for polymer trapping
\be\label{lcr}
L_p^*=\frac{4e^2\sigma_p\sigma_m\ln(d/a)L_m}{\gamma a^2\e_wg\kappa_b^2\Delta P+4\e_w\kappa_b\left(f_p\sigma_p-f_m\sigma_m\right)e\Delta V}.
\ee
Fig.~\ref{fig4}(a) shows that the competition between the barrier and the streaming current results in the decay of the length~(\ref{lcr}) with pressure, i.e. $\Delta P\uparrow L_p^*\downarrow$. As depicted in the same figure, the dilute salt expansion of Eq.~(\ref{lcr}) 
\be\label{lcr2}
L_p^*\approx\frac{4da\ln(d/a)L_m}{(d^2-a^2)\rho_b}\frac{k_BT\sigma_p\sigma_m}{\gamma a^2\Delta P+4(a_p\sigma_p-a_m\sigma_m)e\Delta V}.
\ee
predicts that the same competition leads to the decay of the critical length with added salt, i.e. $\rho_b\uparrow L_p^*\downarrow$.

During polymer capture ($z_p<L_-$), the electrostatic barrier also affects the liquid velocity. For the sake of simplicity, we consider a purely pressure-driven polymer transport and set $\Delta V=0$. The linear PB limit of Eq.~(\ref{vels})  
\be\label{str2}
u_c(r)=\frac{\Delta P}{4\eta L_m}\left[d^2-r^2-2a^2\ln\left(\frac{d}{r}\right)\right]-\frac{\sigma_p\sigma_m}{g\eta\beta\rho_bL_p}\ln\left(\frac{d}{r}\right)
\ee
shows that the barrier slows down the streaming flow around the DNA molecule. This effect is illustrated in Fig.~\ref{fig4}(b). The decrease of the polymer length enhances the barrier and reduces the fluid velocity below the Poiseuille profile (black curve), $\L_p\downarrow u_c(r)\downarrow$. Below the critical length $L_p=L^*_p\approx160$ nm, the velocity of the polymer and the surrounding liquid becomes negative. This prediction suggests that the magnitude of the electrostatic polymer-membrane interactions can be extracted from the streaming current blockade in pressure-driven translocation events.

We finally investigate the effect of pore confinement. Fig.~\ref{fig4}(c) shows that as a result the barrier attenuation, at positive pressures $\Delta P\geq0$, the polymer velocity uniformly rises with the pore radius, $d\uparrow\lan v_p\ran\uparrow$. The reduction of the translocation time with increasing pore radius has been observed in voltage-driven translocation experiments~\cite{e9}. Then, at negative pressures $\Delta P<0$, the velocity initially rises, reaches a peak, and decays at large pore radii ($d\uparrow\lan v_p\ran\downarrow$) where the streaming current opposing the polymer capture overcomes the \textcolor{black}{EP} drift. The cancelation of the polymer velocity at two different pore radii is an observation of practical significance for the design of polymer trapping devices.

\section{Pressure-solvation traps}
\label{cr}

In nanopore-based biosensing approaches, the improvement of the sequencing precision necessitates the mutual enhancement of the capture speed and translocation time~\cite{e5,e6,Tapsarev}. Here, we show that in purely pressure-driven translocation,  this goal can be achieved by adding polyvalent cations to the KCl solution. At vanishing voltage $\Delta V=0$ where the drift velocity~(\ref{vdr}) simplifies to
\be
\label{vdrP}
v_{dr}=\frac{\gamma a^2\Delta P}{4\eta L_m},
\ee
electrostatic interactions come into play only through the interaction potential $V_p(z_p)$ in Eq.~(\ref{polp}). In the presence of polyvalent charges, the derivation of this potential requires the computation of the polymer grand potential $\Delta\Omega_p(l_p)$  beyond MF electrostatics. Sec.~\ref{barcr} reviews the inclusion of the corresponding charge-correlations within the 1l test charge theory developed in Refs.~\cite{the11,the14}.

\subsection{Correlation-corrected grand potential}
\label{barcr}

In the 1l test charge theory, the correlation-corrected polymer grand potential is calculated by approximating the molecule by a charged line located on the pore axis. The corresponding linear charge density is related to the surface charge density of the cylindrical DNA molecule as  $\tau=2\pi a\sigma_{ p}$. The polymer grand potential is obtained by expanding the electrostatic grand potential of charged system at the quadratic order in the polymer charge density $\sigma_p(\br)$ given by Eq.~(\ref{chp}). This expansion yields~\cite{the11}
\be
\label{pr1}
\Delta\Omega_p(l_p)=\Delta\Omega_{mf}(l_p)+\Delta\Omega_s(l_p),
\ee
with the MF component accounting for the direct electrostatic coupling between the polymer and pore charges 
\be\label{grmf2}
\beta\Delta\Omega_{mf}(l_p)=\int\mathrm{d}\br\sigma_p(\br)\phi_m(\br),
\ee
and the polymer self-energy bringing 1l-level electrostatic correlations 
\be
\label{self1}
\beta\Delta\Omega_s(l_p)=\frac{1}{2}\int\mathrm{d}\br\mathrm{d}\br'\sigma_p(\br)\left[v(\br,\br')-v_b(\br-\br')\right]\sigma_p(\br').
\ee

The MF-level grand potential component~(\ref{grmf2}) includes the polymer charge density~(\ref{chp}) and the membrane-induced potential $\phi_m(r)$ solving the PB equation
\be\label{pr2}
\frac{1}{4\pi\ell_Br}\partial_r\left[r\partial_r\phi_m(r)\right]+\sum_{i=1}^3\rho_{bi}q_ie^{-q_i\phi_m(r)}=\sigma_m\delta(r-d).
\ee
Eq.~(\ref{pr2}) cannot be solved in a closed form. The improved Donnan solution of this equation was derived in Ref.~\cite{the14} in the form
\be
\label{pr3}
\phi_m(r)=\phi_d+\frac{4\pi\ell_B\sigma_m}{\kappa_d}
\left[\frac{2}{\kappa_dd}-\frac{\mathrm{I}_0(\kappa_dr)}{\mathrm{I}_1(\kappa_dd)}\right],
\ee
where the Donnan potential $\phi_d$ and screening parameter $\kappa_d$ are obtained from the relations
\be\label{pr4}
\sum_{i=1}^3\rho_{bi}q_ie^{-q_i\phi_d}=\frac{2\sigma_m}{d}\;;\hspace{5mm}\kappa_d^2=4\pi\ell_B\sum_{i=1}^3\rho_{bi}q_i^2e^{-q_i\phi_d}.
\ee
Substituting the potential~(\ref{pr3}) into Eq.~(\ref{grmf2}), one obtains
\be\label{pr5}
\beta\Delta\Omega_{mf}(l_p)=l_p\psi_{mf},
\ee
where we introduced the MF grand potential density
\be\label{pr6}
\psi_{mf}=-\tau\phi_d-\tau\frac{4\pi\ell_B\sigma_m}{\kappa_d}\left[\frac{2}{\kappa_dd}-\frac{1}{\mathrm{I}_1(\kappa_dd)}\right].
\ee

The polymer self-energy ~(\ref{self1}) includes the pore Green's function $v(\br,\br')$ solving the kernel equation
\be\label{7II}
\left[\nabla\e(r)\nabla-\e(r)\kappa^2(r)\right]v(\br,\br')=-\frac{e^2}{k_BT}\delta(\br-\br'),
\ee
with the dielectric permittivity function $\e(r)=\e_w\theta(d-r)+\e_m\theta(r-d)$ and the local screening parameter 
\be\label{kl}
\kappa^2(r)=4\pi\ell_B\sum_{i=1}^3\rho_{bi}q_i^2e^{-q_i\phi_m(r)}\theta(d-r). 
\ee
Eq.~(\ref{self1}) also contains the bulk Green's function $v_b(\br)=\ell_Be^{-\kappa_b|\br|}/|\br|$ where the bulk screening parameter is
\be\label{8}
\kappa_b^2=4\pi\ell_B\sum_{i=1}^3\rho_{bi}q_i^2.
\ee

In Ref.~\cite{the14}, Eq.~(\ref{7II}) was solved within a WKB approach and the self-energy~(\ref{self1}) was obtained in the form
\bea\label{self2}
\beta\Delta\Omega_s(l_p)=l_p\psi_s(l_p),
\eea
with the self-energy per polymer length
\be\label{self2}
\psi_s(l_p)=\ell_B\tau^2\int_{-\infty}^\infty\mathrm{d}k\frac{2\sin^2(kl_p/2)}{\pi l_pk^2}\left\{\ln\left[\frac{p_b}{p(0)}\right]+\frac{Q(k)}{P(k)}\right\}.
\ee
The auxiliary functions in Eq.~(\ref{self2}) are defined as
\bea\label{9}
Q(k)&=&2p^3(d)dB_0(d)\mathrm{K}_0\left(|k|d\right)\mathrm{K}_1\left[B_0(d)\right]\\
&&-2\gamma |k|dp^2(d)B_0(d)\mathrm{K}_1\left(|k|d\right)\mathrm{K}_0\left[B_0(d)\right]\nonumber\\
&&-\left[p^3(d)d-p^2(d)B_0(d)-\kappa(d)\kappa'(d)dB_0(d)\right]\nonumber\\
&&\hspace{3mm}\times\mathrm{K}_0\left(|k|d\right)\mathrm{K}_0\left[B_0(d)\right],\nonumber\\
\label{10}
P(k)&=&2p^3(d)dB_0(d)\mathrm{K}_0\left(|k|d\right)\mathrm{I}_1\left[B_0(d)\right]\\
&&+2\gamma |k|dp^2(d)B_0(d)\mathrm{K}_1\left(|k|d\right)\mathrm{I}_0\left[B_0(d)\right]\nonumber\\
&&+\left[p^3(d)d-p^2(d)B_0(d)-\kappa(d)\kappa'(d)dB_0(d)\right]\nonumber\\
&&\hspace{3mm}\times\mathrm{K}_0\left(|k|d\right)\mathrm{I}_0\left[B_0(d)\right],\nonumber
\eea
with the dielectric contrast parameter $\gamma=\e_m/\e_w$, the screening parameter $p_b=\sqrt{k^2+\kappa_b^2}$, and the functions $p(r)=\sqrt{k^2+\kappa^2(r)}$ and $B_0(r)=\int_0^r\mathrm{d}r'p(r')$. 

In anionic pores characterized by a cation excess,  one has $p(0)>p_b$. Consequently, the logarithmic term of the self energy~(\ref{self2}) is negative. Thus, this attractive \textit{solvation} component favours polymer capture~\cite{rem2}. Then, the second term of Eq.~(\ref{self2}) originating from polymer-image-charge interactions is repulsive and limits polymer penetration. Taking now into account Eq.~(\ref{lpzp}), the polymer-pore interaction potential~(\ref{vp1}) can be finally expressed in terms of the polymer grand potential~(\ref{pr1}) as
\bea\label{potp}
V_p(z_p)&=&\Delta\Omega_p(l_p=z_p)\theta(L_--z_p)\\
&&+\Delta\Omega_p(l_p=L_-)\theta(z_p-L_-)\theta(L_+-z_p)\nonumber\\
&&+\Delta\Omega_p(l_p=L_p+L_m-z_p)\theta(z_p-L_+).\nonumber
\eea

\subsection{Computing translocation time}

In the presence of strong polymer-pore interactions, the drift approximation~(\ref{tau1}) for the polymer translocation time ceases to be accurate. Thus, we derive here the general form of the translocation time. By plugging Eq.~(\ref{vp}) into Eqs.~(\ref{con}) and~(\ref{cur}), the polymer diffusion equation takes the form of a Fokker-Planck equation 
\be
\label{FP}
\partial_tc(z_p,t)=D\partial_{z_p}^2c(z_p,t)+\beta D\partial_{z_p}\left[c(z_p,t)U'_p(z_p)\right]. 
\ee
In the translocation process characterized by Eq.~(\ref{FP}), the mean first passage time $\tau_p(z_2;z_1)$ from the initial point $z_1$ to the final point $z_2$ solves the equation~\cite{muthu} 
\be\label{dyn}
D\partial_{z_1}^2\tau_p(z_2;z_1)-\beta DU'_p(z_1)\partial_{\textcolor{black}{z_1}}\tau_p(z_2;z_1)=-1.
\ee
Solving Eq.~(\ref{dyn}) with reflecting and absorbing boundary conditions respectively at the points $z_1=0$ and $z_2=L_m+L_p$, the translocation time follows as
\be\label{tautot}
\tau_p\equiv\tau_p(L_p+L_m;0)=\tau_c+\tau_d+\tau_e,
\ee
where the capture, pore diffusion, and escape times are respectively
\bea
\label{tauc}
\tau_c&=&I_\tau(0,L_-),\\
\label{taud}
\tau_d&=&I_\tau(L_-,L_+),\\
\label{taue}
\tau_e&=&I_\tau(L_+,L_p+L_m),
\eea
with the auxiliary integral
\be
\label{intt}
I_\tau(z_i,z_f)=\frac{1}{D}\int_{z_i}^{z_f}\mathrm{d}z'e^{\beta U_p(z')}\int_{0}^{z'}\mathrm{d}z''e^{-\beta U_p(z'')}.
\ee

\begin{figure}
\includegraphics[width=1\linewidth]{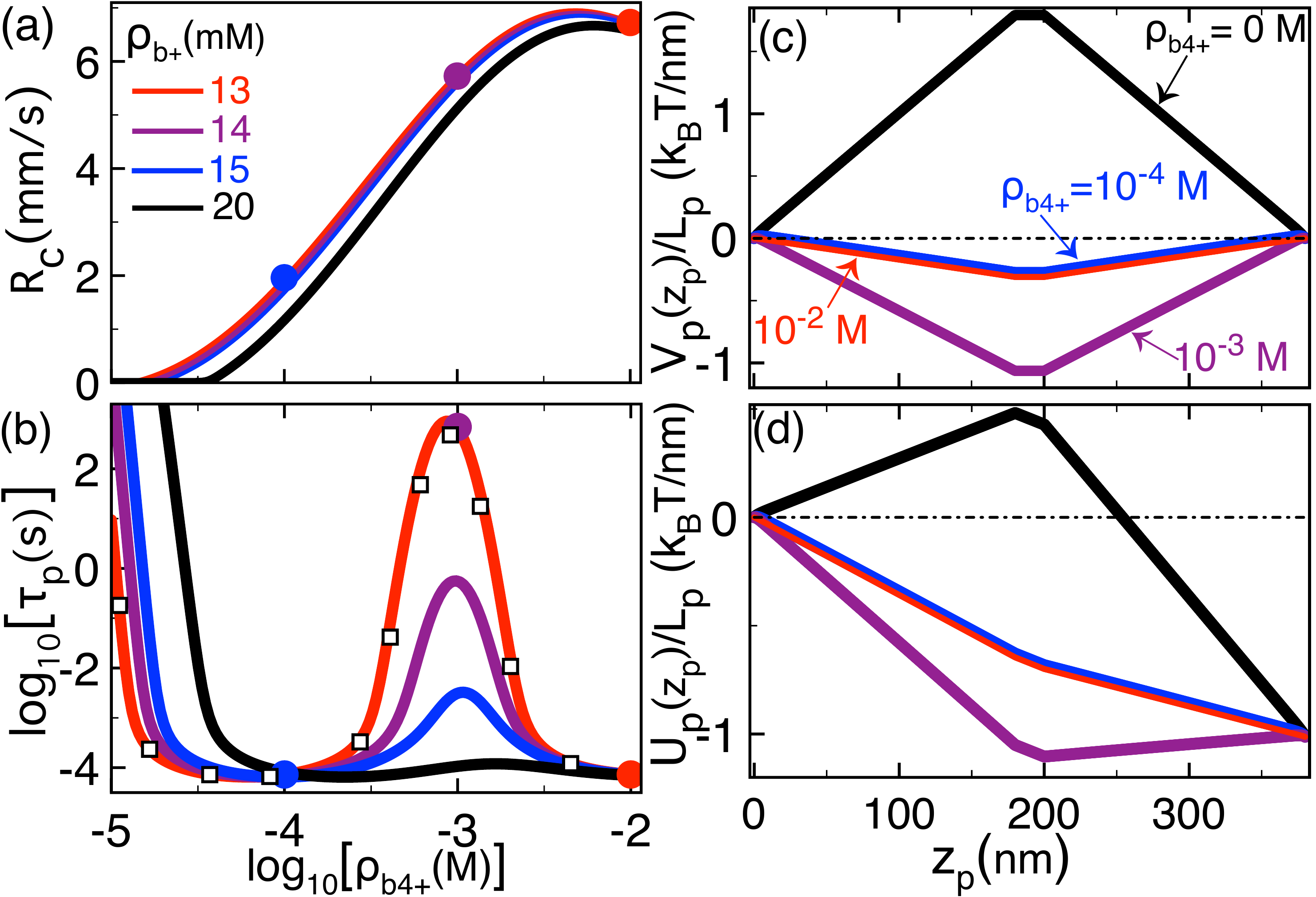}
\caption{(Color online) (a) Translocation rate~(\ref{rc}) and (b) time~(\ref{tautot}) versus the $\mbox{Spm}^{4+}$ density at various KCl densities given in the legend. The open squares in (b) are from Eqs.~(\ref{tc})-(\ref{te}). (c) Interaction potential~(\ref{potp}) and (d) polymer potential~(\ref{polp}) at the monovalent salt density $\rho_{b+}=13$ mM and various $\mbox{Spm}^{4+}$ densities indicated by the dots of the same colour in (a) and (b). The pressure gradient is $\Delta P=2$ atm in all figures. The other parameters are the same as in Fig.~\ref{fig2}.}
\label{fig5}
\end{figure}

\subsection{Faster polymer capture and longer translocation upon $\mbox{Spm}^{4+}$ addition}
\label{mut}

We consider the effect of spermine ($\mbox{Spm}^{4+}$) molecules on polymer capture and translocation. Figs.~\ref{fig5}(a) and (b) illustrate the polymer translocation rates and times versus the $\mbox{Spm}^{4+}$  concentration of the electrolyte $\mbox{KCl}+\mbox{SpmCl}_4$. Figs.~\ref{fig5}(c) and (d) display in turn the polymer-pore interaction and effective potential profiles. In the density regime $\rho_{b4+}\leq10^{-4}$, the addition of $\mbox{Spm}^{4+}$ molecules to the KCl solution enhances the translocation rate and reduces the translocation time, i.e. $\rho_{b4+}\uparrow R_c\uparrow\tau_p\downarrow$. The increase of the translocation speed is induced by the onset of the like-charge polymer-pore attraction; $\mbox{Spm}^{4+}$ molecules screen the repulsive MF-level electrostatic barrier~(\ref{pr6}) and amplify the attractive component of the self-energy~(\ref{self2}). Figs.~\ref{fig5}(c) and (d) show that this switches the interaction potential $V_p(z_p)$ from repulsive to attractive and turns the polymer potential $U_p(z_p)$ to downhill (compare the black and blue curves).

Enhancing further the $\mbox{Spm}^{4+}$ density from $\rho_{b4+}=10^{-4}$ M (blue dots) to $10^{-3}$ M (purple dots), the translocation time rises together with the translocation rate, i.e. $\rho_{b4+}\uparrow R_c\uparrow\tau_p\uparrow$. This intriguing discorrelation between the translocation rate and time originates from the solvation-induced trapping of the polymer. Added $\mbox{Spm}^{4+}$ molecules amplify the like-charge DNA-pore attraction. This enhances the depth of the interaction potential $V_p(z_p)$ and the effective potential $U_p(z_p)$ develops a minimum at $z_p=L_m$ (see the purple curves in Figs.~\ref{fig5}(c) and (d)). Thus, the like-charge DNA-membrane attraction that speeds up the polymer capture also traps the molecule at the pore exit. The consequence of this trapping mechanism on the characteristic times~(\ref{tauc})-(\ref{taue}) is illustrated in Fig.~\ref{fig6}(a). The increment of the $\mbox{Spm}^{4+}$ density from $\rho_{b4+}=10^{-5}$ M to $10^{-3}$ M reduces the polymer capture time and amplifies the escape  time ($\rho_{b4+}\uparrow \tau_c\downarrow\tau_e\uparrow$)  by several orders of magnitude.  This result is the key prediction of our work.

Rising the bulk $\mbox{Spm}^{4+}$ density beyond the value $\rho_{b4+}\approx10^{-3}$ M, charge screening weakens the pore potential $\phi_m(r)$ and the $\mbox{Spm}^{4+}$ excess in the pore. Figs.~\ref{fig5}(c) and (d) show that this attenuates the like-charge DNA-pore attraction and removes the minimum of the effective potential (see the red curves). In Fig.~\ref{fig5}(b), one sees that the removal of the trap at $\rho_{b4+}\gtrsim10^{-3}$ M results in the decrease of the translocation time, i.e.  $\rho_{b4+}\uparrow \tau_p\downarrow$. One also notes that due to the screening of the like-charge attraction, the weak rise of the monovalent salt density reduces the trapping time ($\rho_{b+}\uparrow \tau_p\downarrow$) by orders of magnitude. Thus, the alteration of the monovalent salt density can allow the sensitive tuning of the trapping time.

\subsection{Characterization of the barrier, drift, and trapping regimes}
\label{reg}

In order to gain a quantitative insight into the features discussed in Sec.~\ref{mut}, we evaluate analytically the characteristic times~(\ref{tauc})-(\ref{taue}). To this end, we approximate the self-energy~(\ref{self2}) by its limit reached for a long polymer portion in the pore, i.e. $\kappa_{ b}l_{ p}\gg1$.  This limit reads
\be
\label{selfth}
\lim_{l_p\to\infty}\psi_s(l_p)=\psi_s=\ell_B\tau^2\left\{-\ln\left[\frac{\kappa(0)}{\kappa_{b}}\right]+\frac{Q_0}{P_0}\right\},
\ee
where $Q_0\equiv Q(k\to0)$ and $P_0\equiv P(k\to0)$, or
\bea
\label{q0}
Q_0&=&2\kappa^2(d)dB(d)\mathrm{K}_1\left[B(d)\right]\\
&&-\left\{\kappa^2(d)d-\left[\kappa(d)+\kappa'(d)d\right]B(d)\right\}\mathrm{K}_0\left[B(d)\right],\nonumber\\
P_0&=&2\kappa^2(d)dB(d)\mathrm{I}_1\left[B(d)\right]\\
&&+\left\{\kappa^2(d)d-\left[\kappa(d)+\kappa'(d)d\right]B(d)\right\}\mathrm{I}_0\left[B(d)\right],\nonumber
\eea
with the function $B(r)=\int_0^r\mathrm{d}r'\kappa(r')$. Then, we introduce the characteristic inverse lengths embodying the effect of the drift force and polymer-pore interactions,
\be
\label{lengths}
\ld=\frac{3\pi\beta L_p\gamma a^2\Delta P}{4\ln(L_p/2a) L_m}\;;\hspace{5mm}\lb=\frac{3\ln(d/a)}{2\ln(L_p/2a)}\psi_{tot},
\ee
where we defined the total electrostatic energy density 
\be\label{psitot}
\psi_{tot}=\psi_{mf}+\psi_s, 
\ee
with its MF component $\psi_{mf}$ given by Eq.~(\ref{pr6}). In terms of the inverse lengths~(\ref{lengths}), the polymer potential~(\ref{polp}) takes the piecewise form of Eq.~(\ref{up2}). The characteristic times~(\ref{tauc})-(\ref{taue}) can be now analytically evaluated as
\bea
\label{tc}
\tau_c&=&\frac{1}{D(\ld-\lb)^2}\left[e^{-(\ld-\lb)\lm}-1+(\ld-\lb)\lm\right]\nonumber\\
&&\\
\label{td}
\tau_d&=&\frac{\left[1-e^{-(\ld-\lb)\lm}\right]\left[1-e^{-\ld(\lp-\lm)}\right]}{D\ld(\ld-\lb)}\nonumber\\
&&+\frac{1}{D\ld^2}\left[e^{-\ld(\lp-\lm)}-1+\ld(\lp-\lm)\right]\\
\label{te}
\tau_e&=&\frac{1}{D(\ld+\lb)^2}\left[e^{-(\ld+\lb)\lm}-1+(\ld+\lb)\lm\right]\nonumber\\
&&+\frac{e^{-\ld(\lp-\lm)}}{D(\ld+\lb)}\left[1-e^{-(\ld+\lb)\lm}\right]\\
&&\hspace{3mm}\times\left\{\frac{1-e^{-(\ld-\lb)\lm}}{\ld-\lb}+\frac{1}{\ld}\left[e^{\ld(\lp-\lm)}-1\right]\right\}.\nonumber
\eea
Fig.~\ref{fig5}(b) shows the good accuracy of this approximation (compare the red curve and the square symbols).  
\begin{figure}
\includegraphics[width=.9\linewidth]{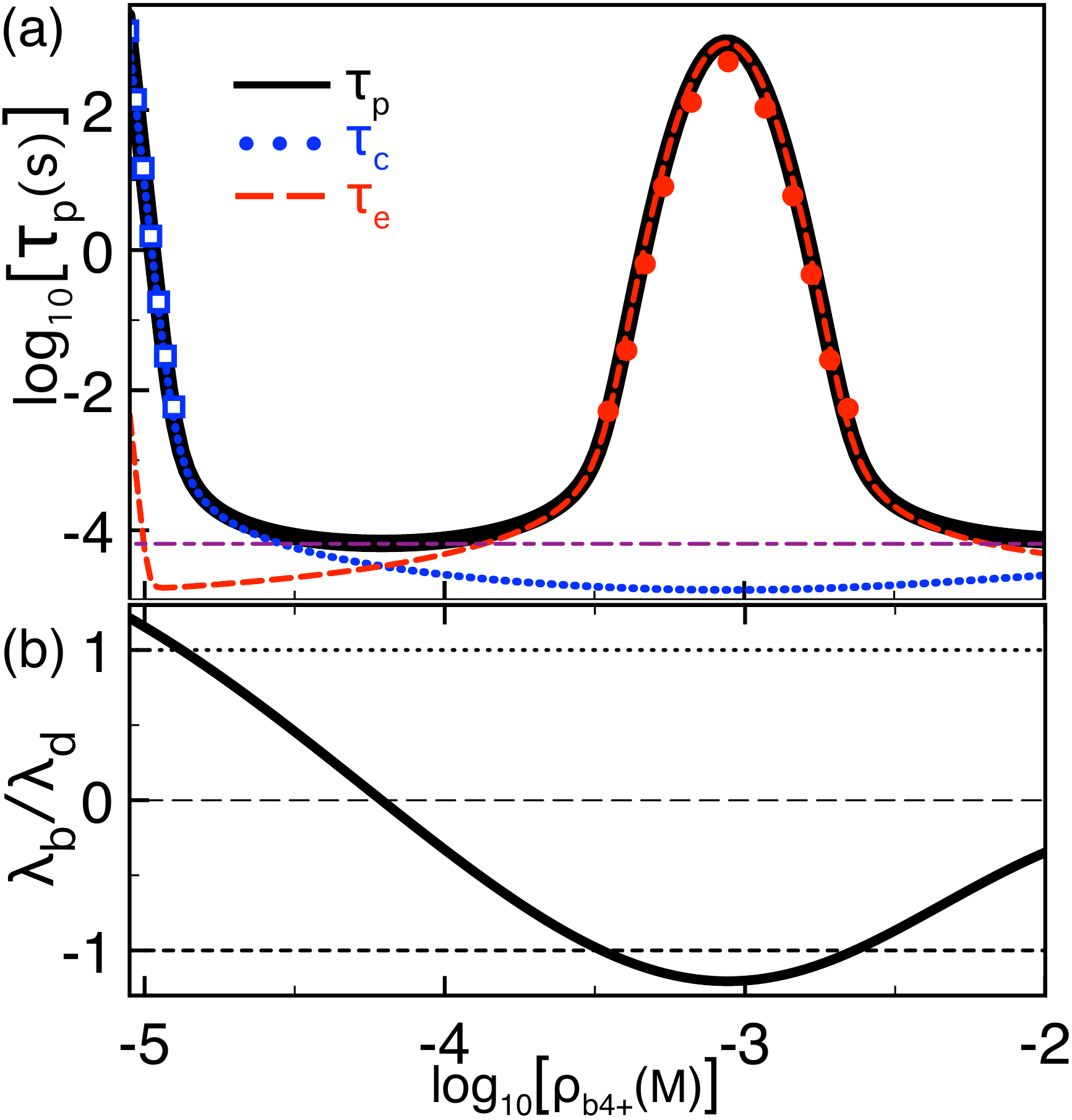}
\caption{(Color online) (a) Translocation time~(\ref{tautot}) (black curve) and its drift limit~(\ref{dron}) (purple curve), capture time~(\ref{tauc}) (blue curve), escape time~(\ref{taue}) (red curve), and (b) the ratio of the  lengths $\ld$ and $\lb$ in Eq.~(\ref{lengths}) against the $\mbox{Spm}^{4+}$ density. In (a), the squares and dots mark respectively the asymptotic laws~(\ref{as1}) and~(\ref{as2}) on their validity regime. The monovalent salt density is $\rho_{b+}=13$ mM and the pressure gradient $\Delta P=2$ atm. The other parameters are the same as in Fig.~\ref{fig2}.}
\label{fig6}
\end{figure}

The effect of $\mbox{Spm}^{4+}$ molecules on the translocation time can be quantitatively characterized in terms of the inverse lengths $\lb$ and $\ld$. Their ratio corresponding to the adimensional interaction potential is displayed in Fig.~\ref{fig6}(b). In the barrier-driven regime $\lb>\ld$ corresponding to the spermine density range $\rho_{b4+}\lesssim10^{-5}$ M, the expansion of Eqs.~(\ref{tc})-(\ref{te}) for $\ld/\lb<1$ yields the characteristic time hierarchy $\tau_c\gg\tau_d\gg\tau_e$ and
\be
\label{as1}
\tau_p\approx\tau_c\approx\frac{e^{(\lb-\ld)L_-}}{D(\lb-\ld)^2}.
\ee
Thus, the capture time is the dominant characteristic time of the barrier-driven regime. The asymptotic law~(\ref{as1}) reported in  Fig.~\ref{fig6}(a) by square symbols corresponds to the Kramer's reaction rate for polymer capture by overcoming the barrier $U_b=k_BT(\lb-\ld)L_-$.

Figs.~\ref{fig6}(a) and (b) show that as one rises the $\mbox{Spm}^{4+}$ density beyond $\rho_{b4+}\approx10^{-5}$ M, the removal of the electrostatic barrier $U_b$ reduces sharply the capture time~(\ref{as1}) and drives the system into the drift-dominated regime $\ld>\lb>-\ld$. Indeed, in the strict limit $|\lb|/\ld\ll1$, the expansion of Eqs.~(\ref{tc})-(\ref{te}) yields the limiting law
\be\label{dron}
\tau_p\approx\tau_{dr}=\frac{L_m+L_p}{v_{dr}}
\ee
indicating purely drift-driven transport at velocity $v_{dr}$. Eq.~(\ref{dron}) is displayed in Fig.~\ref{fig6}(a) by the purple curve.

\begin{figure*}
\includegraphics[width=1\linewidth]{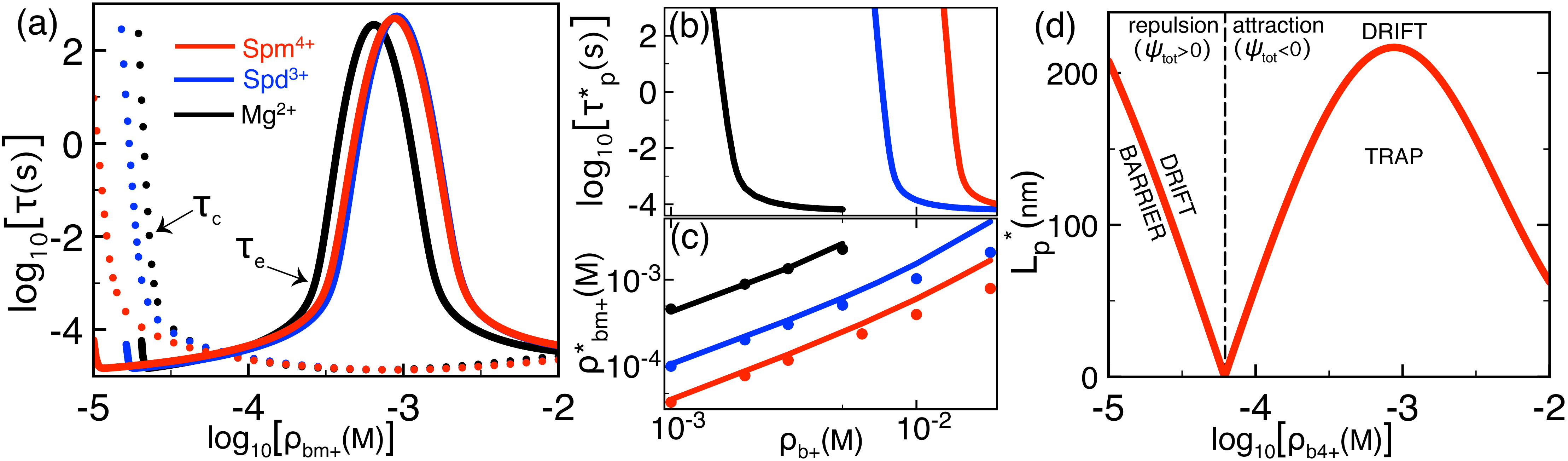}
\caption{(Color online) (a) Polymer capture time~(\ref{tc}) (dots) and escape time~(\ref{te}) (solid curves) against the density of the polyvalent cation species $\mbox{I}^{m+}$  (see the legend) in three different electrolyte mixtures $\mbox{KCl}+\mbox{ICl}_m$. Each mixture has a different $\mbox{K}^+$ density:  $\rho_{b+}=13$ mM ($\mbox{Spm}^{4+}$), $6.8$ mM ($\mbox{Spd}^{3+}$), and $1.5$ mM ($\mbox{Mg}^{2+}$). (b) The peak value $\tau_p^*$ of the translocation time and (c) the corresponding $\mbox{I}^{m+}$ density $\rho_{bm+}^*$ against the bulk $\mbox{K}^+$ concentration. In (c), the dots are from Eq.~(\ref{don3}). (d) Critical polymer length~(\ref{lcr3}) splitting the barrier, drift, and trapping regimes in the $\mbox{Spm}^{4+}$ liquid. The pressure gradient is $\Delta P=2$ atm. The other parameters are the same as in Fig.~\ref{fig2}.}
\label{fig7}
\end{figure*}

In Fig.~\ref{fig6}(b), one sees that the increase of the $\mbox{Spm}^{4+}$ density further beyond the value $\rho_{b4+}\approx10^{-3.5}$ M drives the sytem into the trapping regime $\lb<-\ld$. Expanding Eqs.~(\ref{tc})-(\ref{te}) for $\lb/\ld<-1$, one gets $\tau_e\gg\tau_{c,d}$ and
\be
\label{as2}
\tau_p\approx\tau_e\approx\frac{-\lb}{D\ld(\lb+\ld)^2}e^{-(\lb+\ld)L_-}.
\ee
Hence, in the trapping regime, the escape time dominates the translocation. The asymptotic law~(\ref{as2}) displayed in Fig.~\ref{fig6}(a) by circles corresponds to the reaction rate for the unbinding of the polymer from the pore exit where the molecule is trapped in a potential well of depth $U_b=k_BT|\lb+\ld|L_-$.  In this regime, the abrupt rise of the escape time~(\ref{as2}) upon $\mbox{Spm}^{4+}$ addition stems precisely from the lowering of the trap depth $U_b$ by the intensification of the like-charge polymer-pore attraction. 

At this point, the question arises whether the solvation-induced trapping can be induced by counterions of lower valency. Fig.~\ref{fig7}(a) displays the polymer capture and escape times in three different electrolyte mixtures $\mbox{KCl}+\mbox{ICl}_m$.  Each solution has a different bulk $\mbox{K}^+$ density indicated in the caption. The figure shows that as long as the monovalent salt concentration of the liquid is lowered together with the valency of the multivalent cation species $\mbox{I}^{m+}$, trivalent $\mbox{Spd}^{3+}$ and divalent $\mbox{Mg}^{2+}$ counterions can reduce the capture time and extend the escape time as efficiently as quadrivalent $\mbox{Spm}^{4+}$ molecules. In Fig.~\ref{fig7}(b), this point is illustrated in terms of the peak translocation time versus the monovalent salt density. One notes that the lower the valency of the polyvalent counterion species, the lower the $\mbox{K}^+$ density range where the maximum translocation time rises sharply.

In Fig.~\ref{fig6}, the correlation between $\tau_p$ and $\lb/\ld$  indicates that the polyvalent cation density $\rho^*_{bm}$ maximizing the trapping time  can be evaluated by identifying the minimum of the grand potential~(\ref{psitot}). To this end, we pass to the pure Donnan approximation and set $\phi(r)\to\phi_d$. The screening function~(\ref{kl}) becomes $\kappa(r)=\kappa_d$. Consequently, the grand potential density~(\ref{psitot}) simplifies to
\be
\label{don1}
\psi_{tot}\approx-\tau\phi_d+\ell_B\tau^2\left\{-\ln\left(\frac{\kappa_d}{\kappa_{b}}\right)+\frac{\mathrm{K}_1(\kappa_dd)}{\mathrm{I}_1(\kappa_dd)}\right\}.
\ee
To progress further, we consider the Gouy-Chapman (GC) regime of dilute salt $\kappa_b\mu\ll1$ with the GC length $\mu=1/(2\pi\ell_B\sigma_m)$. Expanding the equalities in Eq.~(\ref{pr4}), at leading order, the Debye potential and screening parameter follow as $\phi_d\approx-\ln\left[2\sigma_m/(m\rho_{bm+}d)\right]/m$ and $\kappa_d^2\approx8\pi\ell_Bm\sigma_m/d$. Substituting these equalities into Eq.~(\ref{don1}) and carrying out another expansion for $\kappa_b\mu\ll1$, the grand potential density finally becomes
\bea
\label{don2}
\psi_{tot}&\approx&\frac{\tau}{m}\ln\left(\frac{2\sigma_m}{m\rho_{bm+}d}\right)\\
&&-\frac{\ell_B\tau^2}{2}\ln\left\{\frac{2m\sigma_m}{d\left[2\rho_{b+}+(m^2+m)\rho_{bm+}\right]}\right\}.\nonumber
\eea
The density $\rho_{bm}^*$ maximizing the translocation time $\tau_p$ follows from the equation $\partial\psi_{tot}/\partial\rho_{bm+}=0$  as
\be
\label{don3}
\rho_{bm+}^*=\frac{4\rho_{b+}}{(m^2+m)\left(m\ell_B\tau-2\right)}.
\ee
In the derivation of the density~(\ref{don3}), the system was assumed to be in the trapping regime. This requires both the polymer self-energy and the grand potential~(\ref{don2}) to be negative. Thus, the polymer charge density should satisfy the inequality $\tau>2/(m\ell_B)$.  Fig.~\ref{fig7}(c) illustrates the numerically evaluated characteristic density $\rho_{bm+}^*$ (solid curves) together with the analytical estimation~(\ref{don3}) (dots). Eq.~(\ref{don3}) indicates that $\rho^*_{bm}$  rises linearly with the $\mbox{K}^+$ concentration ($\rho_{b+}\uparrow\rho^*_{bm}\uparrow$) and drops rapidly with the polyvalent counterion valency according to an inverse cubic polynomial law ($m\uparrow\rho^*_{bm}\downarrow$).

Finally, we characterize finite-size effects on polymer trapping. By equating the characteristic inverse lengths in Eq.~(\ref{lengths}), the critical polymer length separating the drift and interaction-dominated regimes follows as
\be
\label{lcr3}
L_p^*=\frac{2\ln(d/a)L_m}{\pi\beta\gamma a^2\Delta P}\left|\psi_{tot}\right|
\ee
Fig.~\ref{fig7}(d) displays Eq.~(\ref{lcr3}) against the $\mbox{Spm}^{4+}$ density. The transition from the drift-driven ($L_p>L^*_p$) to the barrier/trapping regime ($L_p<L^*_p$) upon polymer length reduction stems from the decrease of the pressure-induced drag force on the polymer. The corresponding balance between polymer-pore interactions and the drift force was scrutinized in Section~\ref{fin} for monovalent solutions.

In the dilute $\mbox{Spm}^{4+}$ regime of Fig.~\ref{fig7}(d) characterized by repulsive polymer-pore interactions ($\psi_{tot}>0$), added $\mbox{Spm}^{4+}$ molecules suppress the electrostatic barrier and lower the critical length, i.e. $\rho_{b4+}\uparrow\left|\psi_{tot}\right|\downarrow L_p^*\downarrow$.  In the subsequent $\mbox{Spm}^{4+}$ density range where the like-charge polymer-pore attraction is activated ($\psi_{tot}<0$), $\mbox{Spm}^{4+}$ addition enhances the trapping potential depth and rises the critical polymer length, $\rho_{b4+}\uparrow\left|\psi_{tot}\right|\uparrow L_p^*\uparrow$. Beyond the density value $\rho_{b4+}\approx1$  mM, added $\mbox{Spm}^{4+}$ molecules screen the attractive polymer-pore interactions. This reduces the depth of the potential trap and drops the critical length. To conclude, polymer trapping by like-charge attraction occurs if the polymer length satisfies the condition $L_p<L_p^*$. The upper polymer length~(\ref{lcr3}) can be however tuned by controlling the mangitude of the potential $\psi_{tot}$ via the alteration of the ion density.

\section{Conclusions}

The optimization of polymer translocation techniques requires the accurate characterization of the electrohydrodynamic forces governing driven polymer transport. In this article, we characterized the collective effect of the \textcolor{black}{EP} drift, the drag force induced by the streaming flow, and electrostatic polymer-pore interactions on polymer translocation through solid-state pores. Our main results are summarized below. 

In the first part, we investigated the polymer conductivity of pressure-voltage traps in monovalent salt solutions. By direct comparison with experimental data, we showed that our theory can accurately reproduce and explain the pressure dependence of the polymer translocation velocity and time. Then, we characterized the effect of salt density variation.  In translocation events driven by streaming flow ($\Delta P>0$) and limited by voltage ($\Delta V<0$), added salt screens the negative \textcolor{black}{EP} mobility and favours polymer capture. In the opposite case of voltage-driven ($\Delta V>0$) and pressure-limited translocation ($\Delta P<0$), the polymer mobility exhibits a non-monotonical salt dependence; dilute salt screens electrostatic polymer-pore interactions and favours polymer capture but strong salt reduces the \textcolor{black}{EP} mobility and blocks polymer transport. This non-uniform behavior results in the trapping of the polymer at two distinct salt density values given by Eqs.~(\ref{dil2}) and~(\ref{dil4}). 

We also found that during  polymer capture, the repulsive polymer-pore coupling can reduce or even invert the direction of the streaming current.  Due to the amplification of the barrier effect, the reduction of the liquid velocity becomes stronger with decreasing polymer length. This suggests that electrostatic polymer-pore interactions can be probed by streaming current measurements carried-out at different polymer lengths.

The precision of polymer sequencing by translocation is known to depend on the fast capture of the polymer by a like-charged pore followed by a slow translocation. In the second part of our work, we identified an electrostatic polymer trapping mechanism that allows to achieve this condition by the simple addition of polyvalent cations to the KCl solution. Enhanced electrostatic correlations upon $\mbox{Spm}^{4+}$ addition turn the polymer-pore interactions from repulsive to attractive. This like-charge polymer-pore attraction results in a faster polymer capture from the cis side but traps the molecule at the pore exit on the trans side of the membrane. As a result, the increment of the $\mbox{Spm}^{4+}$ density from $\rho_{b4+}=10^{-5}$ M to $10^{-3}$ M reduces the capture time and extends the escape time ($\rho_{b4+}\uparrow\tau_c\downarrow\tau_e\uparrow$) by five orders of magnitude. 

Provided that the monovalent salt density is lowered together with the valency of the polyvalent counterions, trivalent $\mbox{Spd}^{3+}$ and divalent $\mbox{Mg}^{2+}$ cations can trap the polymer as efficiently as quadrivalent $\mbox{Spm}^{4+}$ molecules. Eq.~(\ref{don3}) indicates that the polyvalent ion density $\rho_{bm+}^*$ minimizing the capture time and maximizing the trapping time rises with the monovalent salt concentration $\rho_{b+}\uparrow\rho_{bm+}^*\uparrow$ and drops with the ionic valency $m\uparrow\rho_{bm+}^*\downarrow$. Finally, we showed that solvation-induced polymer trapping can be achieved only if the molecular length is below the critical length $L_p^*$ given by Eq.~(\ref{lcr3}). It should be noted that the maximum length $L_p^*$ can be tuned by the alteration of the ion density.

Our formalism neglects some features of these highly complex systems, such as conformational polymer fluctuations~\cite{Duncan1999}, entropic barriers limiting polymer capture~\cite{n2}, the discrete charge distribution on the membrane surface and the helicoidal charge partition on the polymer~\cite{Sung2013}. \textcolor{black}{Our translocation model does not include either the interaction of the membrane with the polymer portion outside the pore}, as well as hydrodynamic and electrostatic edge effects occuring at the pore ends~\cite{Levin2006}. Although the consequence of these approximations cannot be estimated quantitatively without the explicit inclusion of the corresponding effects, the agreement with experimental data indicates that in the experimental configuration considered herein, these complications play a secondary role. \textcolor{black}{For example, as discussed in Section~\ref{mod}, the accuracy of the stiff polymer approximation is due to the short length of the DNA sequences involved in the translocation experiments of Ref.~\cite{exp2}.} It should be also noted that in the low pressure regime of Fig.~\ref{fig2} where the net drift force on DNA becomes rather weak, entropic effects expected to become relevant may be responsible for the slight deviation of our theoretical curves from the experimental trend. \textcolor{black}{In order to understand the electrohydrodynamics of translocation for long polymer sequences, at the first step, we plan to include to our model the interaction of the membrane matrix with the polymer portion outside the pore. At the next step, the inclusion of conformational polymer fluctuations will allow to take into account the tension propagation mechanism introduced by Sakaue~\cite{Saka1,Saka2,Saka3}.} We finally note that our \textcolor{black}{results and conclusions} can be corroborated by current polymer transport experiments. In particular, the polyvalent cation-induced trapping can be easily verified by standard pressure-driven translocation experiments carried-out with anionic nanopores. Our \textcolor{black}{numerious predictions} can also guide the optimized conception of new generation biosensing tools.
\\
\acknowledgements  This work was performed as part of the Academy of Finland Centre of Excellence program (project 312298).

\smallskip
\appendix
\section{Coefficients of the average polymer velocity formula~(\ref{bv})}
\label{coef}

We list here the coefficients of the average velocity formula~(\ref{bv}) of the main text, 

\bea
J_1&=&\frac{1}{\left(\ld-\lb\right)^2}\left\{\left(\ld-\lb\right)L_-+e^{-(\ld-\lb)L_-}-1\right\}\nonumber\\
&&+\frac{1}{\ld-\lb}\left[1-e^{-(\ld-\lb)L_-}\right]\\
&&\times\left\{\frac{1}{\ld}\left[1-e^{-\ld(L_+-L_-)}\right]\right.\nonumber\\
&&\hspace{5mm}\left.+\frac{1}{\ld+\lb}e^{-\ld(L_+-L_-)}\left[1-e^{-(\ld+\lb)L_-}\right]\right\}\nonumber\\
J_2&=&\frac{1}{\ld^2}\left\{\ld(L_+-L_-)+e^{-\ld(L_+-L_-)}-1\right\}\\
&&+\frac{1-e^{-(\ld+\lb)L_-}}{\ld(\ld+\lb)}\left[1-e^{-\ld(L_+-L_-)}\right]\nonumber\\
J_3&=&\frac{1}{\left(\ld+\lb\right)^2}\left\{\left(\ld+\lb\right)L_-+e^{-(\ld+\lb)L_-}-1\right\}.\nonumber\\
\eea


\begin{thebibliography}{99}
\bibitem {rem1} The parameter $L_m$ corresponds to the position of the effective pore boundary where the polymer is removed from the system on the trans side. Complications associated with the electrohydrodynamic edge effects at the pore ends, the deviation of the pore geometry from a perfect cylinder, and the variation of the electric field outside the pore are adsorbed into the effective parameter $L_m$.
\end{thebibliography}
\end{document}